\documentclass[letterpaper,prd,10pt,twocolumn,showpacs,showkeys,preprintnumbers,superscriptaddress,amsmath,amssymb,nofootinbib,notitlepage]{revtex4-1}
\usepackage{graphicx}
\usepackage{amsmath}
\usepackage{amsfonts}
\usepackage{amsthm}
\usepackage{amssymb}
\usepackage{graphicx}
\usepackage{epstopdf}
\usepackage{hyperref}

\usepackage{subfig}

	\newcommand{\ncd}{\newcommand}
	\ncd{\mrm}    {\mathrm}
	\ncd{\beq} {\begin{equation}}
	\ncd{\eeq} {\end{equation}}

	\def\d{{\rm d}}
	\def\D{{\rm D}}

	\newtheorem{defn}{Definition}

\begin{document}
\title{The geometry of electromagnetic curves on Riemannian manifolds}

\author{Sergio Islas-Ram\'irez,}
\email[]{al2202803149@azc.uam.mx}
\affiliation{Universidad Aut\'onoma Metropolitana Azcapotzalco,  Avenida San Pablo Xalpa 180, Azcapotzalco, Reynosa Tamaulipas, C.P. 02200, Ciudad de M\'exico, Mexico}

\author{Cesar S. Lopez-Monsalvo}
\email[]{cslopezmo@conacyt.mx}
\affiliation{Conacyt-Universidad Aut\'onoma Metropolitana Azcapotzalco,  Avenida San Pablo Xalpa 180, Azcapotzalco, Reynosa Tamaulipas, C.P. 02200, Ciudad de M\'exico, Mexico}

\author{Jos\'e Antonio Eduardo Roa-Neri}
\email[]{rnjae@azc.uam.mx}
\affiliation{Universidad Aut\'onoma Metropolitana Azcapotzalco,  Avenida San Pablo Xalpa 180, Azcapotzalco, Reynosa Tamaulipas, C.P. 02200, Ciudad de M\'exico, Mexico}

%

\begin{abstract}
We present a concise definition of an electromagnetic curve on a Riemannian manifold and illustrate the explicit case of  the motion of a charged particle on the unit sphere under the influence of a uniform magnetic field.
\end{abstract}

\maketitle
\section{Introduction}

Magnetic curves describe the motion of a charged particle under the influence of a magnetic field. That is, solutions to the equations of motion
	\beq
	\frac{\d \vec p}{\d t} = q \left(\vec v \times \vec B \right) \quad {\rm with} \quad \vec p = m \vec v.
	\eeq
Here, $\vec p$ is the momentum of the particle, $q$ and $m$ represent its charge and mass, respectively, $\vec v$ is its velocity and $\vec B$ is a given magnetic field \cite{barros2005gauss}.

On the other hand, Riemannian geometry has proven to be extremely useful in describing the dynamics of a system  subject to spatial constraints \cite{arnol2013mathematical}. That is, situations in which the entire space is not available and the motion is confined to  a given surface. In such case, a reformulation of the Principle of Inertia is in order, providing us with the conditions a curve must satisfy so that it describes free motion on the constraint surface. 

In this work, we provide a concise definition of electromagnetic curves on general Riemannian manifolds aided with the geometric formulation of Maxwell's theory \cite{lopez2020geometry}, allowing us to present the dynamics for an arbitrary number of spatial dimensions, where the tools of vector calculus are no longer well defined.

\begin{figure}
\centerline{
\subfloat[Magnetic curve on free space]{\includegraphics[width=0.5\columnwidth]{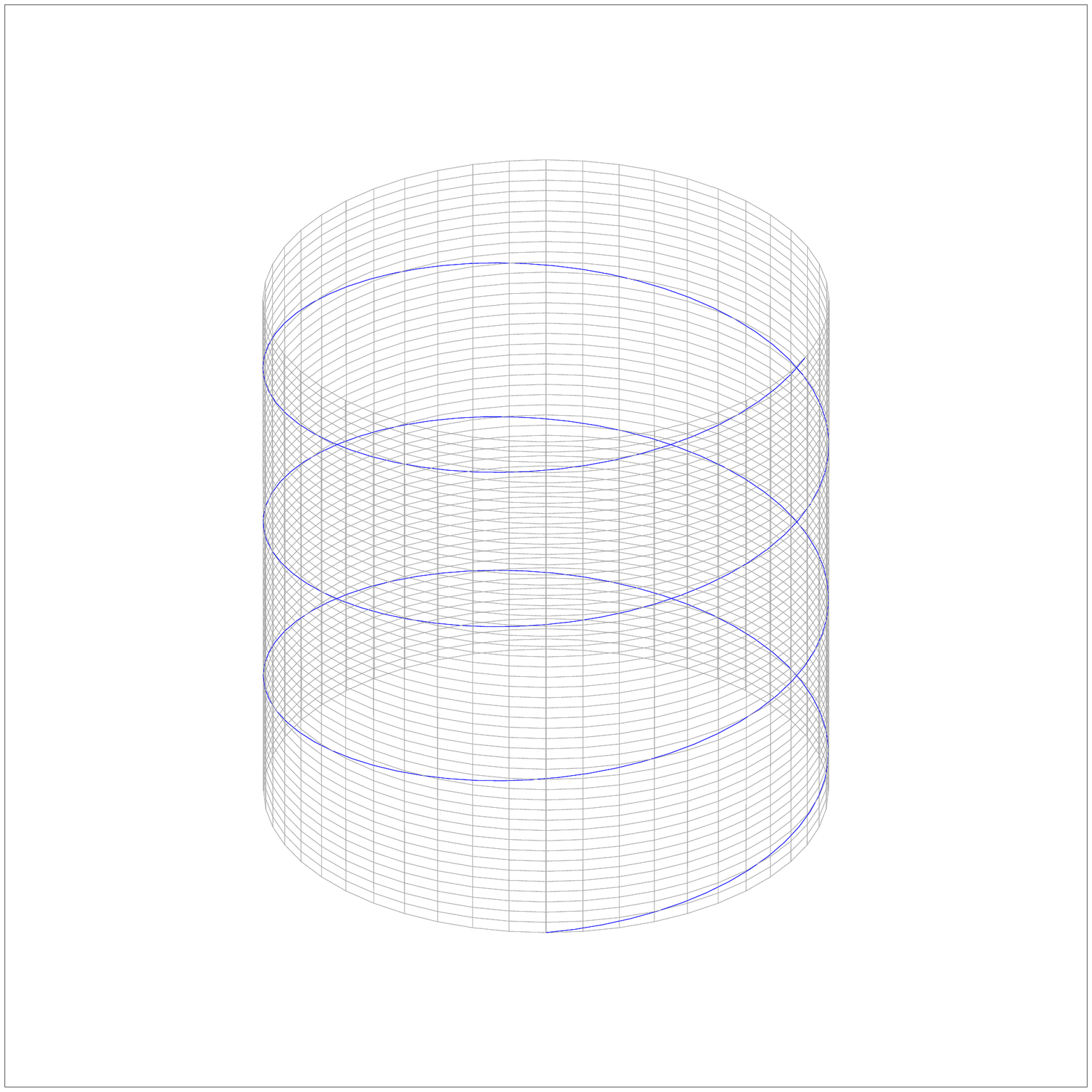}}
\hfill
\subfloat[Free motion on a sphere]{\includegraphics[width=0.5\columnwidth]{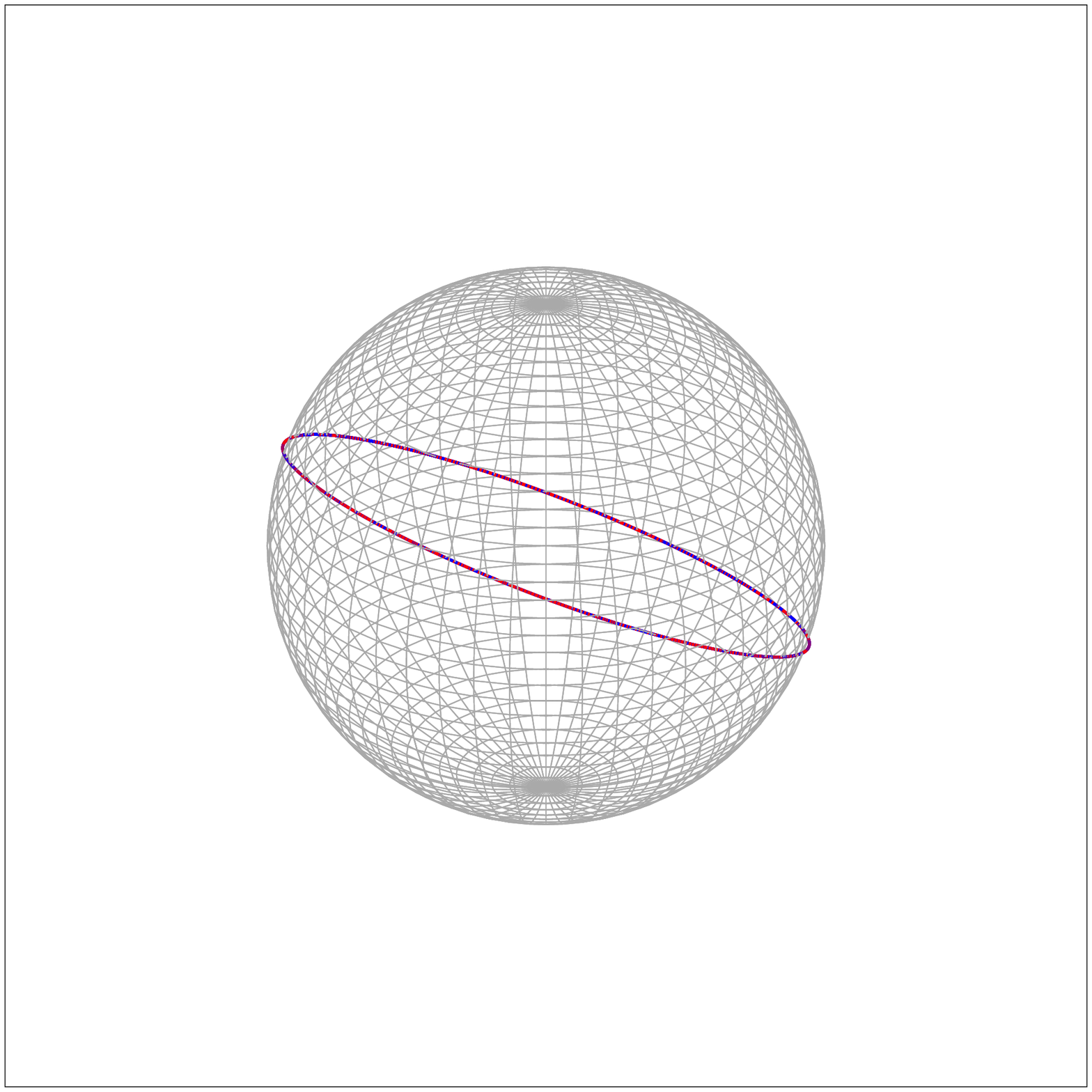}}}
\caption{Magnetic curve for a uniform magnetic field in free space and a curve of minimal length on a sphere.}
\end{figure}

\section{Inertial motion on a Riemannian manifold}
%
 
 Galilean inertia -- rooted in Euclidean geometry -- is a degenerate case of differential geometry in the sense that straight lines, \emph{auto-parallel} curves and paths of minimal length, coincide. Soon after Lobachevsky and Bolyai ventured outside the realm of Euclid's axioms, new possibilities for inertial motion emerged. In particular, one can formulate a Riemannian Principle of Inertia, namely
 \begin{defn}[Riemannian Inertial Motion] Let $(M,g)$ be a Riemannian manifold where $g$ denotes its metric. Consider a parametrized curve
		\beq
		\gamma:[a,b]\subset \mathbb{R} \rightarrow M.
		\eeq
	
We say $\gamma$ is an inertial motion if:
	\begin{enumerate}
	 \item It is an extremal of the arc-length functional
	 	\beq
		\label{eq.arc}
		\ell(\gamma) = \int_a^b \sqrt{g(\dot \gamma,\dot \gamma)}\ \d t \quad {\rm{where}} \quad \dot \gamma \equiv \frac{\d}{\d t} \gamma 
		\eeq
	\item Its velocity is uniform, i.e.
		\beq
		\label{eq.unif}
		\frac{\d}{\d t} \vert\dot\gamma \vert^2 = \pounds_{\dot \gamma} g(\dot \gamma,\dot \gamma) = 0 \quad \forall t \in [a,b],
		\eeq
	where $\pounds$ denotes the \emph{Lie} derivative \cite{nakahara2003geometry}.
	\end{enumerate}
	\end{defn}

It follows from \eqref{eq.arc} and \eqref{eq.unif} that acceleration measures the departure of a curve from being inertial. This is expressed through the notion of \emph{covariant derivative} which, in the case of Riemannian geometry,  is given by 
 	\beq
	\label{eq.accel}
		a \equiv \nabla_{\dot \gamma} {\dot \gamma} = \frac{\D}{\d t} \dot \gamma = \frac{\D}{\d t} \frac{\d}{\d t} \gamma,
		\eeq
	where $\nabla$ is the Levi-Civita connection compatible with the metric $g$ and $\D/\d t$ denotes its associated covariant derivative \cite{carmo1992riemannian}.
 
\section{Electromagnetic curves on Riemannian manifolds}

Maxwell's equations on manifolds are expressed as two independent conservation laws \cite{lopez2020geometry}, namely
	\beq
	\label{eq.maxwell}
	\oint_{\partial \Omega^3} F \overset{!}{=} 0 \quad {\rm and} \quad \oint_{\partial \Omega^n} J \overset{!}{=} 0.
	\eeq
Here $\Omega^3$ and $\Omega^n$ represent arbitrary three and $n$ dimensional regions with boundary, respectively. The boundary operator is expressed as $\partial$ and the symbol $\overset{!}{=}$ denotes a \emph{physical} demand, in this case, that the electromagnetic flux $F$ and current $J$ are conserved.  Stokes' theorem 
together with the arbitrariness of the domains in \eqref{eq.maxwell} imply the local conservation laws
	\beq
	\d F = 0 \quad {\rm and } \quad \d J = 0,
	\eeq
where d denotes the \emph{exterior derivative}. Therefore, an electromagnetic field on a manifold is expressed by a closed 2-form $F$ whilst an electromagnetic current is given by a closed $(n-1)$-form $J$. 

On a Riemannian manifold $(M,g)$, the metric tensor plays the r$\hat {\rm o}$le of a material medium, defining  the constitutive relation
	\beq
	\label{eq.maxwell2}
	H = \zeta \star_g F \quad {\rm with} \quad \d H = J,
	\eeq
where $\zeta$ denotes the medium impedance and $\star_g$ is the \emph{Hodge star operator} associated with the metric $g$ \cite{nakahara2003geometry}.

The motion of a charged particle under the influence of an electromagnetic field $F$ satisfies the equation 
	\beq
	\label{eq.lorentz}
	m \nabla_{\dot \gamma} \dot \gamma = q\Phi(\dot \gamma),
	\eeq
where the left hand side (lhs) is the particle's mass times its acceleration  [cf. equation \eqref{eq.accel}] while the right hand side (rhs) is the Lorentz force defined by the compatibility condition	
	\beq
	\label{eq.comp}
	g\left(\Phi(u),w\right) = F(u,w),
	\eeq
where $u$ and $w$ are two arbitrary vector fields defined on $M$ \cite{barros2005gauss}.

Therefore, an  electromagnetic curve on a Riemannian manifod can be defined as follows:
	\begin{defn}[Electromagnetic curve]
	Let $(M,g)$ be a Riemannian manifold, $J$ a closed $(n-1)$-form on $M$. A parametrized curve
		\beq
		\gamma:[a,b]\subset\mathbb{R}\rightarrow M
		\eeq 
		is called an electromagnetic curve if it satisfies \eqref{eq.lorentz} where the electromagnetic field $F$ is a solution to Maxwell's equations \eqref{eq.maxwell2} and the Lorentz force satisfies the compatibility condition \eqref{eq.comp}.
		\end{defn}

	
\section{Magnetic curves on a sphere}

\begin{figure}
\centerline{
\subfloat[Front]{\includegraphics[width=0.5\columnwidth]{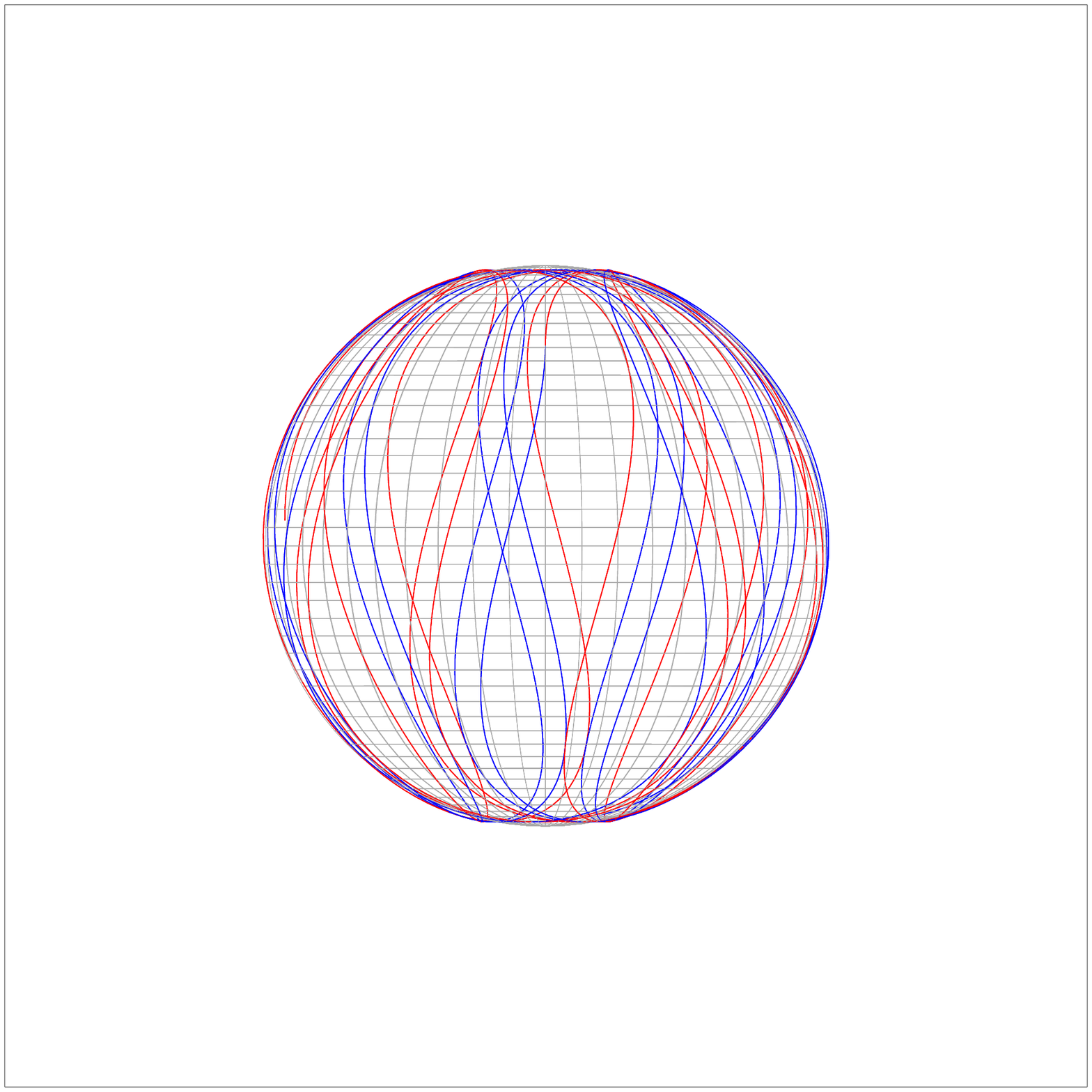}}
\hfill
\subfloat[Top]{\includegraphics[width=0.5\columnwidth]{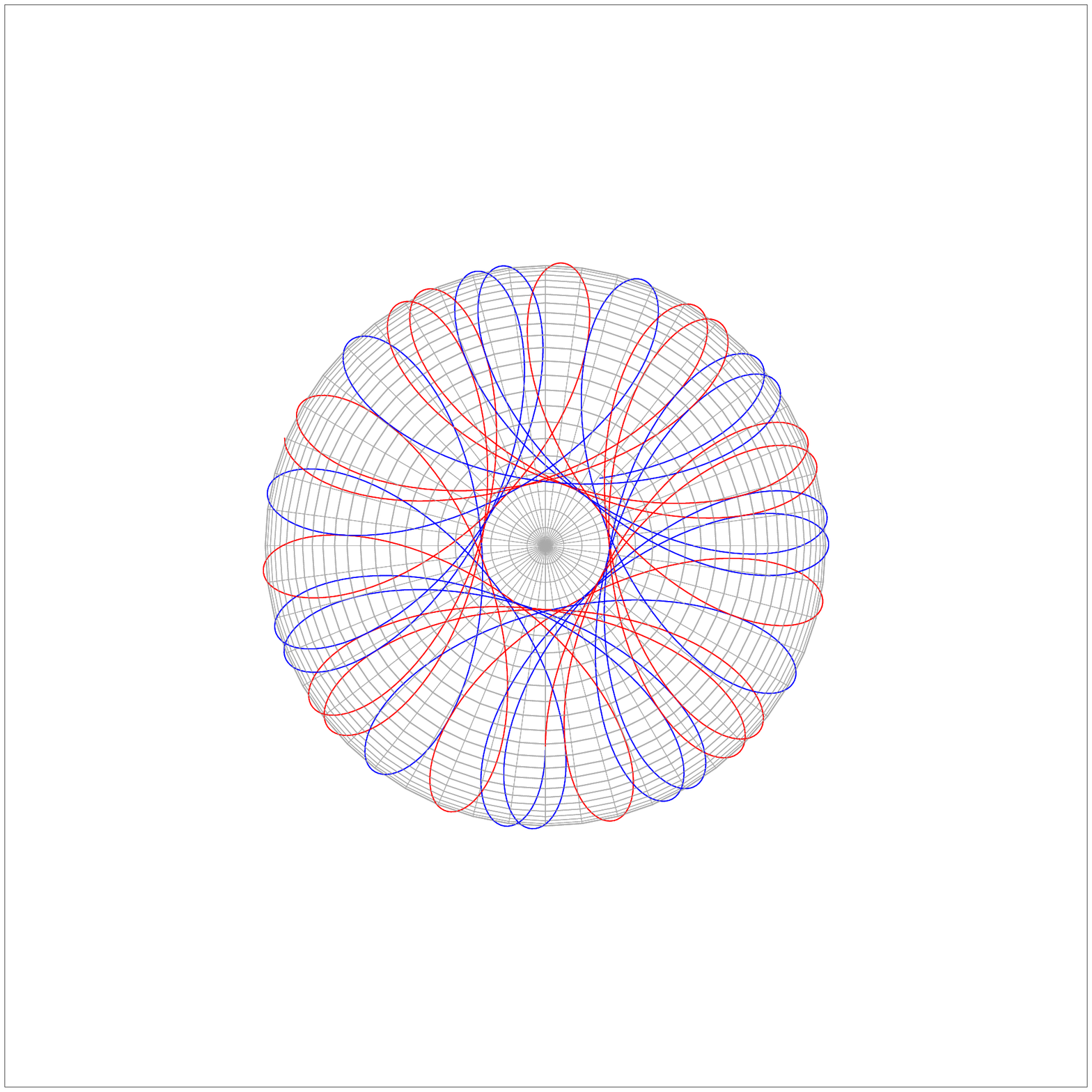}}}

\centerline{
\subfloat[Front]{\includegraphics[width=0.5\columnwidth]{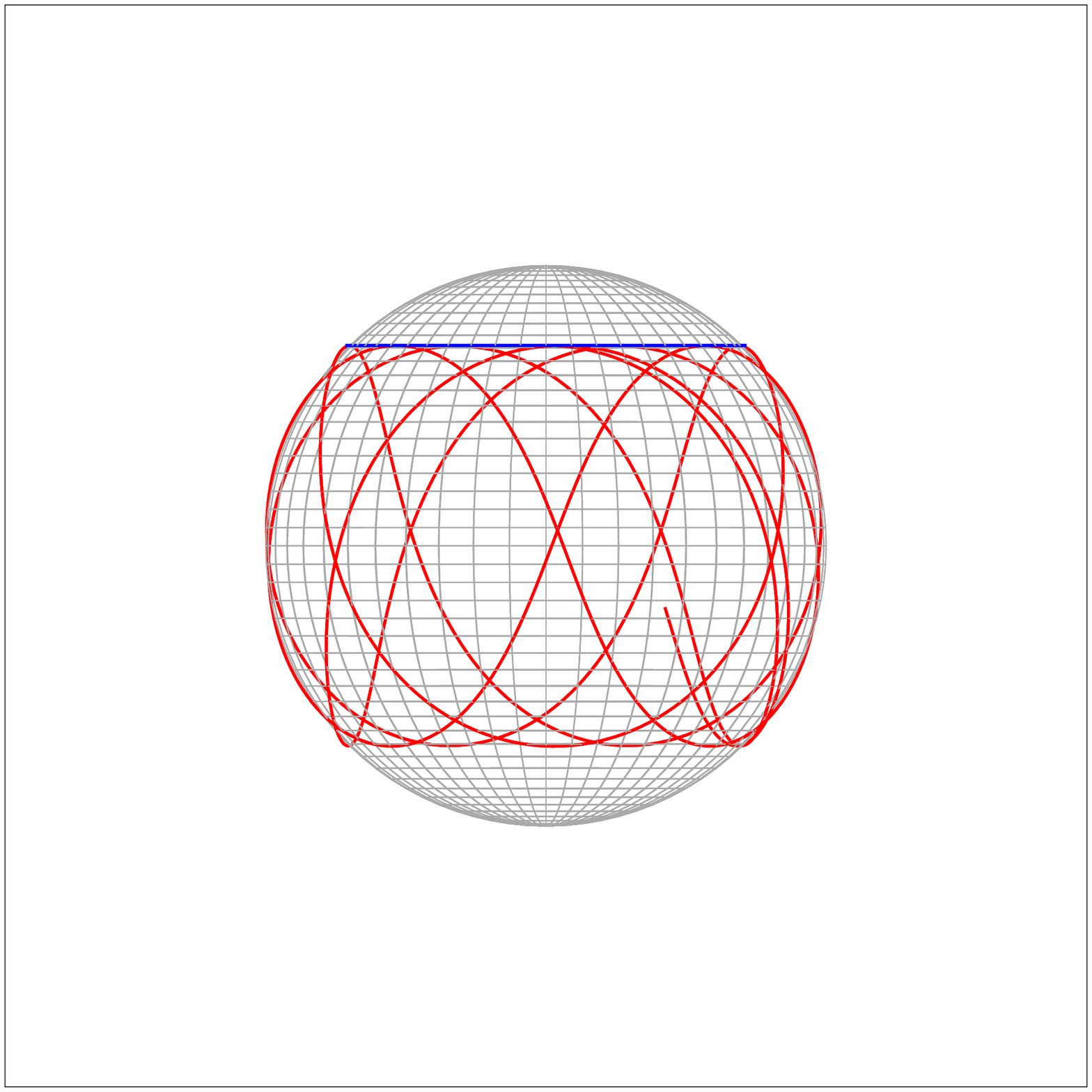}}
\hfill
\subfloat[Top]{\includegraphics[width=0.5\columnwidth]{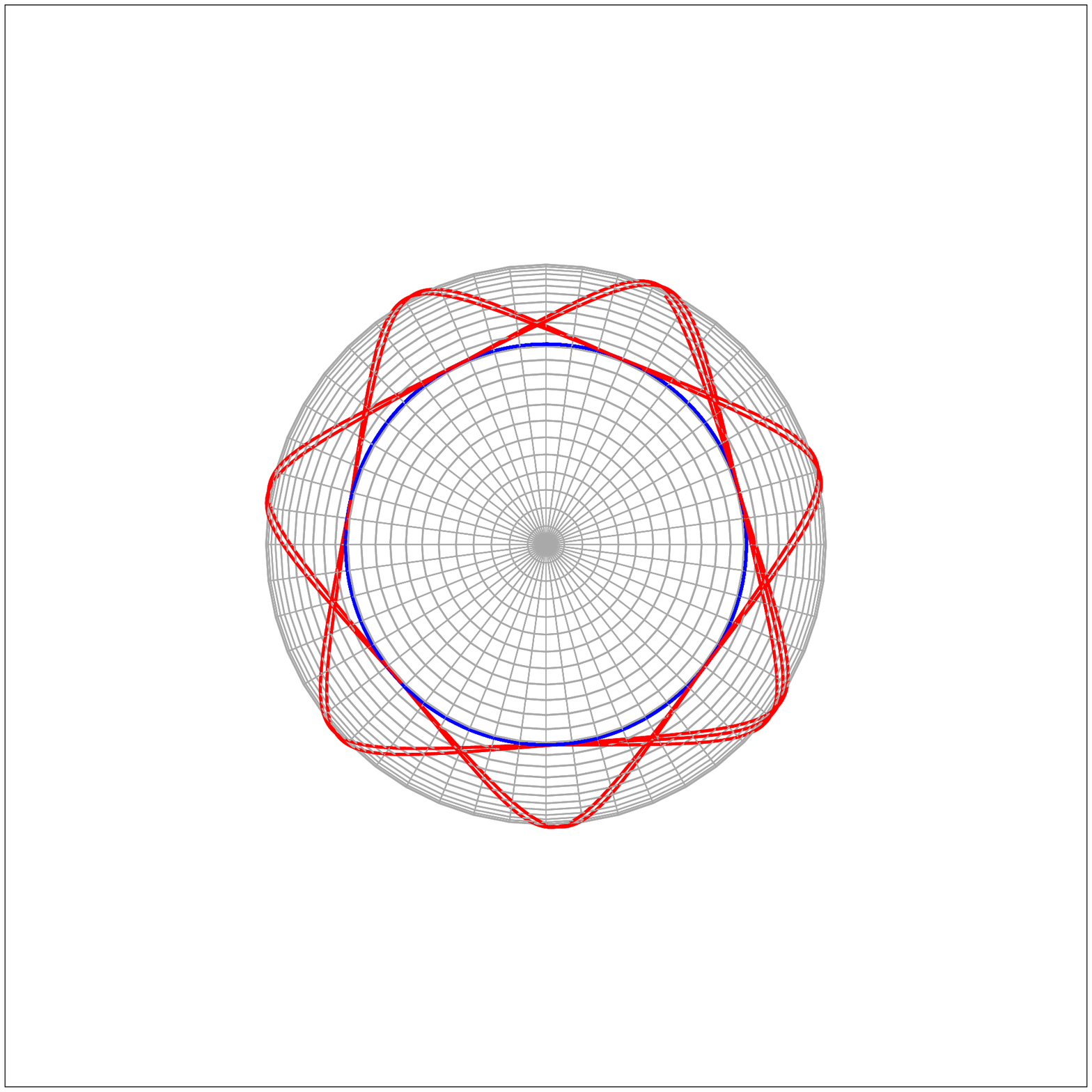}}
}
\caption{Magnetic curves on a sphere. The upper figures correspond to initial conditions along the longitudinal direction while the lower ones along the parallel lines. }\label{fig.2}
\end{figure}

Let us consider a vertically oriented uniform magnetic field defined in a region of $\mathbb{R}^3$ where the standard Euclidean inner product is assumed, 
 together with a charged particle confined to move on a sphere. Let $(S^2,g)$ be the unit sphere   canonically embedded in $\mathbb{R}^3$  where $g$ is the induced  metric given by
	\beq
	g = \d \theta \otimes \d \theta + \sin^2(\theta) \d \varphi \otimes \d \varphi
	\eeq
and the flux 2-form on the sphere becomes
	\beq
	F\vert_{S^2} = B_z \cos(\theta) \sin(\theta) \ \d \theta \wedge \d \varphi.
	\eeq
	
The particle's acceleration  \eqref{eq.accel} in the lhs of \eqref{eq.lorentz} is given by
	\begin{align}
	a = \left[\frac{\d^2 \theta}{\d t^2} + \sin(\theta) \cos(\theta) \left(\frac{\d\varphi}{\d t}\right)^2\right] \frac{\partial}{\partial \theta}\nonumber\\
	 +\left[\frac{\d^2\varphi}{\d t^2} + \cot(\theta) \frac{\d \theta}{\d t}\frac{\d \varphi}{\d t} \right]\frac{\partial}{\partial \varphi},
	\end{align}
while 
	\begin{align}
	\Phi(\dot \gamma) = \left[ B_z \sin(\theta)\cos(\theta)\frac{\d \varphi}{\d t}\right] \frac{\partial }{\partial \theta}\nonumber\\
		-\left[ B_z \cot(\theta)\frac{\d\theta}{\d t} \right]\frac{\partial}{\partial \varphi}.
	\end{align}

Solving numerically equation \eqref{eq.lorentz} for $\gamma:t\mapsto [\theta(t),\varphi(t)]$ we obtain the curves shown in figure \ref{fig.2}. Here, we present some representative cases of magnetic curves for a pair of initial conditions. The red and blue lines correspond to flipped initial velocities. Notice the directionality of the curves, they are not reversible. This fact motivates further exploration and provides us with a magnetic analogue to the \emph{Zermelo navigation problem} \cite{bao2004zermelo}.

\bibliographystyle{IEEEtran}
\bibliography{myrefs}

\end{document}